\documentclass[aps,prl,preprint,superscriptaddress,endfloats,showpacs]{revtex4}

\usepackage[utf8]{inputenc}
\usepackage{amsmath}
\usepackage{graphicx}
\usepackage{epstopdf}
\usepackage{color}
\usepackage{tabularx}

\begin{document}

\title{Diffraction based Hanbury Brown and Twiss interferometry performed at a hard x-ray free-electron laser}
\date{\today}

\author{O. Yu.~Gorobtsov}
	\affiliation{Deutsches Elektronen-Synchrotron DESY, Notkestra{\ss}e 85, D-22607 Hamburg, Germany}
\author{N.~Mukharamova}
	\affiliation{Deutsches Elektronen-Synchrotron DESY, Notkestra{\ss}e 85, D-22607 Hamburg, Germany}
\author{S.~Lazarev}
	\affiliation{Deutsches Elektronen-Synchrotron DESY, Notkestra{\ss}e 85, D-22607 Hamburg, Germany}
	\affiliation{National Research Tomsk Polytechnic University (TPU), pr. Lenina 30, 634050 Tomsk, Russia}
\author{M. Chollet}
	\affiliation{SLAC National Accelerator Laboratory, 2575 Sand Hill Rd., Menlo Park, 94025 CA}
\author{D. Zhu}
	\affiliation{SLAC National Accelerator Laboratory, 2575 Sand Hill Rd., Menlo Park, 94025 CA}
\author{Y. Feng}
	\affiliation{SLAC National Accelerator Laboratory, 2575 Sand Hill Rd., Menlo Park, 94025 CA}
\author{R.P.~Kurta}
	\affiliation{Deutsches Elektronen-Synchrotron DESY, Notkestra{\ss}e 85, D-22607 Hamburg, Germany}
	\affiliation{Present address: European XFEL GmbH, Holzkoppel 4, D-22869 Schenefeld, Germany}	
\author{J.-M. Meijer}
        \affiliation{Van 't Hoff Laboratory for Physical and Colloid Chemistry, Debye Institute for Nanomaterial Science, Utrecht University, Padualaan 8, 3584 CH Utrecht, The Netherlands}
        \affiliation{Present address: Division of Physical Chemistry, Department of Chemistry, Lund University, 22100 Lund, Sweden.}
\author{G. Williams}
	\affiliation{SLAC National Accelerator Laboratory, 2575 Sand Hill Rd., Menlo Park, 94025 CA}
	\affiliation{Present address: U.S. Department of Energy Brookhaven National Laboratory, 53 Bell Avenue, Upton, NY 11973-5000, USA}
\author{M. Sikorski}
	\affiliation{SLAC National Accelerator Laboratory, 2575 Sand Hill Rd., Menlo Park, 94025 CA}
	\affiliation{Present address: European XFEL GmbH, Holzkoppel 4, D-22869 Schenefeld, Germany}
\author{S. Song}
	\affiliation{SLAC National Accelerator Laboratory, 2575 Sand Hill Rd., Menlo Park, 94025 CA}
\author{D.~Dzhigaev}
	\affiliation{Deutsches Elektronen-Synchrotron DESY, Notkestra{\ss}e 85, D-22607 Hamburg, Germany}
\author{S.~Serkez}
	\affiliation{European XFEL GmbH, Holzkoppel 4, D-22869 Schenefeld, Germany}	
\author{A.~Singer}
	\affiliation{University of California San Diego, 9500 Gilman Dr., La Jolla, California 92093, USA}
\author{A.V. Petukhov}
    \affiliation{Van 't Hoff Laboratory for Physical and Colloid Chemistry, Debye Institute for Nanomaterial Science, Utrecht University, Padualaan 8, 3584 CH Utrecht, The Netherlands}
    \affiliation{Laboratory of Physical Chemistry, Department of Chemical Engineering and Chemistry, Eindhoven University of Technology, P.O. Box 513, 5600 MB Eindhoven, Netherlands}
\author{I. A. ~Vartanyants}
\email[Corresponding author: ]{ivan.vartaniants@desy.de}
	\affiliation{Deutsches Elektronen-Synchrotron DESY, Notkestra{\ss}e 85, D-22607 Hamburg, Germany}
	\affiliation{National Research Nuclear University MEPhI (Moscow Engineering Physics Institute), Kashirskoe shosse 31, 115409 Moscow,
    Russia}

\begin{abstract}
We demonstrate experimentally Hanbury Brown and Twiss (HBT) interferometry at hard X-ray Free Electron Laser (XFEL) on sample diffraction patterns.
This is different from the traditional approach when HBT interferometry requires direct beam measurements in absence of the sample.
HBT analysis was carried out on Bragg peaks from the colloidal crystals measured at Linac Coherent Light Source (LCLS).
We observed nearly perfect (90\%) spatial coherence and the pulse duration on the order of 11 fs for the monochromatized beam that is significantly shorter than expected from the electron bunch measurements.
\end{abstract}


\maketitle

X-ray free-electron lasers (XFELs) provide extremely bright and highly coherent x-ray radiation with femtosecond pulse duration.
They find extensive applications in the wide range of scientific fields: structural biology \cite{Seibert2011, Chapman2011}, solid density plasma \cite{Vinko2012}, matter under extreme conditions \cite{Schropp2015}, ultrafast photochemistry \cite{LiekhusSchmaltz2015}, atomic physics \cite{Young2010} and many others.
XFEL coherence properties often significantly affect its experimental performance.
Several methods were employed to study spatial and temporal coherence, such as double pinholes \cite{Vartanyants2011, Singer2012}, Michelson type interferometry \cite{Schlotter2010, Roling2011, Singer2012, Hilbert2014}, speckle contrast analysis \cite{Gutt2012, Lee2013, Lehmkuhler2014}, and Hanbury Brown and Twiss (HBT) interferometry \cite{Singer2013, Singer2013_Erratum, Song2014, Gorobtsov2017} (see for review \cite{Vartanyants2016}).

One important aspect that makes XFELs substantially different from all other existing x-ray sources, is the degeneracy parameter, or average number of x-ray photons in one state.
If for present high-brilliance synchrotron sources this value is about $10^{-2}$, for the XFEL sources it can reach such high values as $10^{10}$~\cite{Singer2012, Singer2013, Singer2013_Erratum, Stohr2017}.
This makes XFEL sources similar to optical lasers, and implies possibility of non-linear and quantum optics experiments, as was first suggested by Glauber~\cite{Glauber1963}.
This area in the FEL science is just on its early stage of development~\cite{Rohringer2012, Tamasaku2014, Wu2016, Rudenko2017}.
At the core of the quantum optics experiments stays HBT interferometry \cite{Hanbury1956_1, Hanbury1956_2}.
Since its first demonstration it was used, for example, to analyze nuclear scattering experiments \cite{Baym1998}, to probe Bose-Einstein condensates \cite{Schellekens2005,  Folling2005, Polkovnikov2006} or to study effects of interaction on HBT interferometry \cite{Bromberg2010}.
HBT interferometry is especially well suited to study statistical behavior of XFELs due to their extremely short pulse duration.
It allows to extract both the spatial and temporal XFEL coherence properties \cite{Singer2012, Singer2013, Singer2013_Erratum, Gorobtsov2017} as well as statistical information about the secondary beams and positional jitter \cite{Gorobtsov2017}.

The basic idea of the HBT interferometry \cite{Hanbury1956_1, Hanbury1956_2} is to determine statistical properties of radiation from the normalized second-order intensity-intensity correlation function
\begin{equation}
g^{(2)}(\mathbf{r_1}, \mathbf{r_2}) = \frac{\langle I(\mathbf{r_1}) I(\mathbf{r_2}) \rangle}{\langle I(\mathbf{r_1})\rangle \langle I(\mathbf{r_2}) \rangle}
\label{coherence::2_order_normalized}
\end{equation}
obtained by measuring the coincident response of two detectors at separated positions $\mathbf{r_1}$ and $\mathbf{r_2}$
(see for review~\cite{Paul1986}).
In Eq. \eqref{coherence::2_order_normalized}, $I(\mathbf{r_1})$, $I(\mathbf{r_2})$ are the intensities of the wave field and
the averaging denoted by brackets $<...>$ is performed over a large ensemble of different realizations of the wave field, or different pulses in the case of XFEL radiation.

In this Letter we present results of HBT interferometry performed on the Bragg peaks originating from the scattering on colloidal crystals.
Due to a small beam size and large sample-detector distance instead of the conventional sharp Bragg peaks a comparably broad intensity distribution around each Bragg peak position is measured.
Importantly, this intensity distribution depends not only on the crystal structure, but also on the incident pulse profile.
Statistical changes of the XFEL beam structure from pulse to pulse lead to corresponding changes in the observed Bragg peaks intensity distribution.
Therefore, fluctuating behavior of the Bragg peak intensity contains information about the statistical properties of the incident radiation typical for self-amplified spontaneous emission (SASE) XFELs \cite{SaldinBook}.
This allowed us to extract information on statistical properties of the XFEL radiation during diffraction experiment on colloidal crystals.

\begin{figure}
        \includegraphics[width=\linewidth]{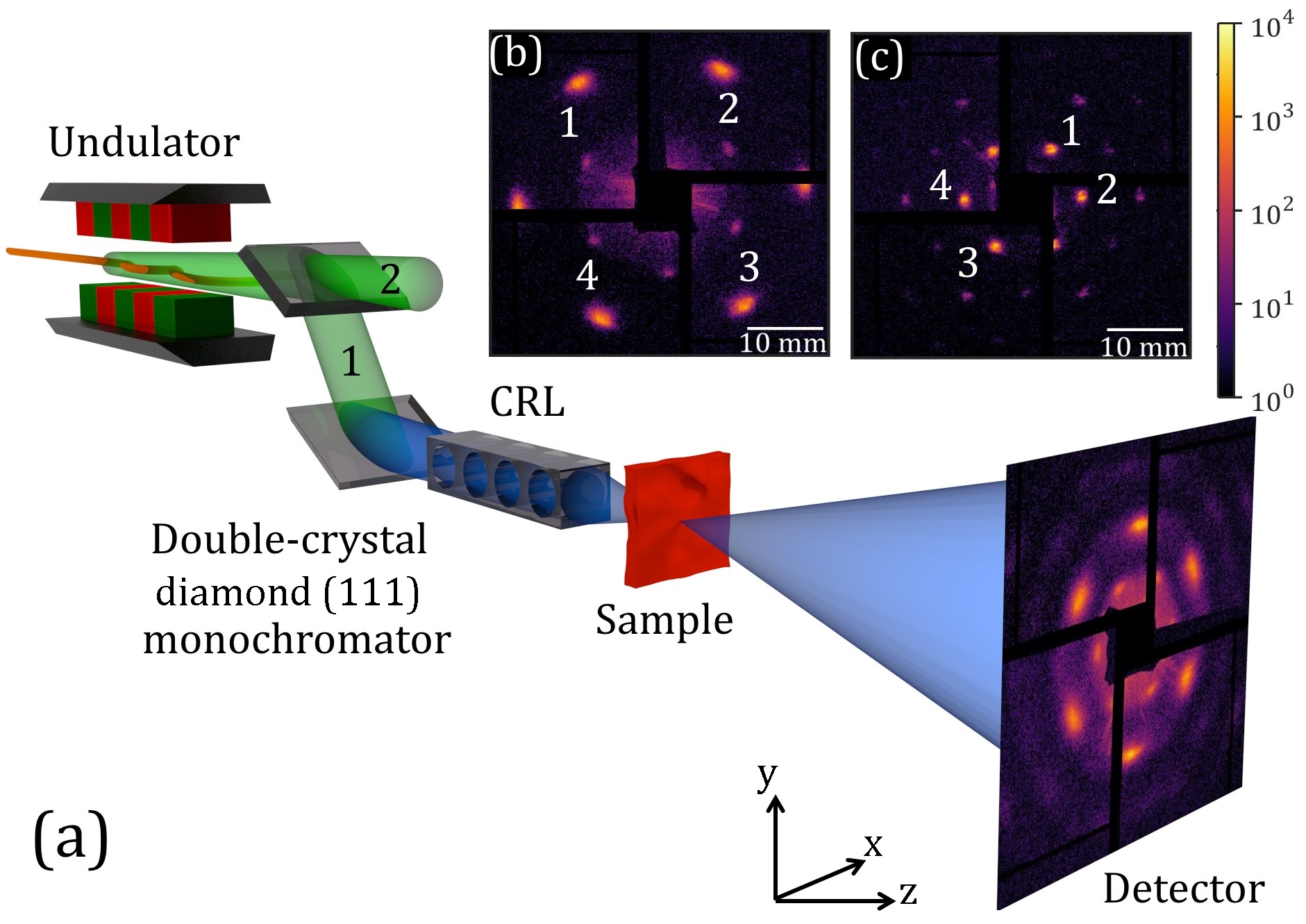}
        \caption{\label{fig::Scheme}
                (Color online)
                (a) Schematic layout of the experiment.
                LCLS radiation passes a double-crystal diamond (111) monocromator and is separated into diffracted (1) and transmitted (2) branches.
                The monochromatized radiation in the diffracted branch is focused on the sample by the compound refractive lenses (CRLs).
                Diffracted intensities are measured by the CSPAD detector positioned $10$ m downstream from the sample.
                Central part of the typical diffraction patterns (shown in the log-scale) for a PS crystal with $160$ nm sphere size (sample 1) (b) and a PS crystal with $420$ nm sphere size (sample 2) (c).
                The peaks chosen for analysis are marked with white numbers.
        }
        \newpage
\end{figure}

\begin{figure}[h]
        \includegraphics[width=\linewidth]{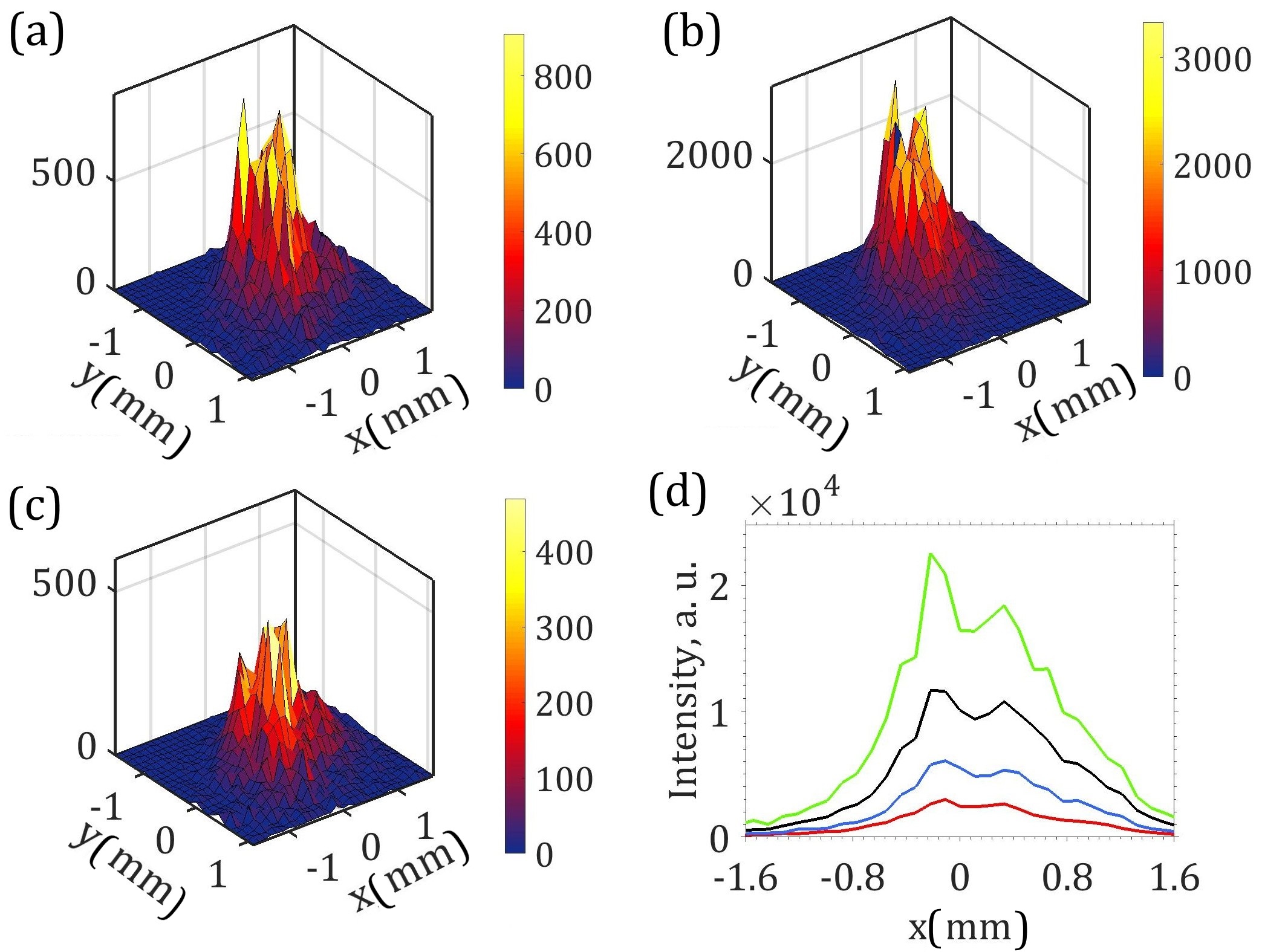}
        \caption{\label{fig::osc}
                (Color online)
                Single pulse intensities measured at the Bragg peak 4 for sample 2.
                (a-c) Typical 2D Bragg peak intensity distribution for different incoming pulses.
                (d) Projection of these intensities on the horizontal direction (intensity shown in (a) - blue curve, in (b) - green curve, and in (c) - red curve) and an average projected intensity for $50,000$ pulses (black).
        }
        \newpage
\end{figure}

The measurements were performed at Linac Coherent Light Source (LCLS) in Stanford, USA at the x-ray pump probe (XPP) beamline \cite{Chollet2015}.
LCLS was tuned to produce pulses with $3.3-3.7$ mJ pulse energy, bunch charge $0.18$ nC, and pulse repetition rate $120$ Hz.
An expected pulse duration from electron bunch measurements was about $41-43$ fs.
The double-crystal diamond (111) monochromator at LCLS with the thicknesses of the monochromator crystals 100 $\mu$m and 300 $\mu$m split the primary x-ray beam into a pink (transmitted) and monochromatic (diffracted) branches (see Fig.~\ref{fig::Scheme}(a)).
We used the monochromatic branch with the photon energy of 8 keV (1.55 \AA) and relative energy bandwidth of $4.4 \cdot 10^{-5}$ \cite{Zhu2014}.
Compound refractive lenses (CRLs) focused the beam at the sample position down to 50 $\mu$m full width at half maximum (FWHM).
The number of photons in the focus was about $10^9$ ph/pulse, and the experiment was performed in non-destructive mode
\footnote{This was confirmed by comparing diffraction patterns in the beginning and the end of the run.}.
Colloidal crystal sample was positioned vertically, perpendicular to the incoming XFEL pulse in the transmission diffraction geometry (see Fig.~\ref{fig::Scheme}(a)).
Series of x-ray diffraction patterns were recorded using the CSPAD megapixel x-ray detector positioned at the distance $L = 10$ m downstream from the sample consisting of 32 silicon sensors with pixel size of 110 x 110 $\mu$m$^2$ covering an area of approximately 17 x 17 cm$^2$ (see for experimental details Ref. \cite{Mukharamova2017}).

Colloidal crystal films were prepared from the polystyrene (PS) using the vertical deposition method~\cite{JM_Meijer_Book}.
The film consisted of 30-40 monolayers of spherical particles.
Two samples were investigated: PS colloidal crystals with a sphere diameter of $160 \pm 3$ nm (sample 1) and $420 \pm 9$ nm (sample 2).

Examples of diffraction patterns measured from these crystals are shown in Fig.~\ref{fig::Scheme} (b) and (c).
Bragg peaks with the six-fold symmetry originating from the hexagonal colloidal crystal structure are clearly visible in this figure~\cite{Sulyanova2015}.
Intensity distributions around the Bragg peak 4 for sample 2 for three different incident pulses at the same position of the sample are shown in Fig.~\ref{fig::osc}.
It is well seen from this figure that Bragg peak profiles for each pulse have a complicated internal structure with additional sub-peaks.
These sub-peaks have the same position from pulse to pulse but their relative intensity varies.
Projection on the horizontal axis of the same Bragg peak intensities for three selected pulses as well as an average projected intensity for all pulses is shown in Fig.~\ref{fig::osc}(d).

In our experimental geometry we were in Fresnel scattering conditions (Fresnel number $1.7$).
It can be shown (see Appendix for details) that in general case of Fresnel scattering the intensity-intensity correlation function at a selected Bragg peak is given by an expression
\begin{equation}
g^{(2)}(\mathbf{Q}, \mathbf{Q'})= 1 + \zeta_2(\sigma_{\omega})\left| \mu(\mathbf{Q}, \mathbf{Q'})\right|^2 \ .
\label{supp::g2_final_spectr}
\end{equation}
Here vector $\mathbf{Q}$ is related to a radius vector $\mathbf{r}$, measured from the center of the diffraction peak, by the relation $\mathbf{Q}=k\mathbf{r}/L$, where $k=2\pi/\lambda$ and $\lambda$ is the wavelength
\footnote{In the following we perform evaluation in $\mathbf{r}$-space.}.
The contrast function $\zeta_2(\sigma_{\omega})$ introduced in Eq.~\eqref{supp::g2_final_spectr} is strongly dependent on the radiation bandwidth $\sigma_\omega$.
The spectral degree of coherence $\mu(\mathbf{Q}, \mathbf{Q'})$ in Eq.~\eqref{supp::g2_final_spectr} is defined as
$\mu(\mathbf{Q}, \mathbf{Q'}) = J(\mathbf{Q}, \mathbf{Q'})/\sqrt{\langle I(\mathbf{Q}) \rangle}\sqrt{\langle I(\mathbf{Q'}) \rangle}$,
where $J(\mathbf{Q}, \mathbf{Q'})$ is the mutual intensity function (MIF) determined at the detector position.
It is directly related to the statistical properties of the incident beam at the sample position by a two-dimensional Fourier transform
\begin{equation}
	\left| J(\mathbf{Q}, \mathbf{Q'}) \right|  = \left|\iint e^{-i(\mathbf{Q'}\mathbf{r'}-\mathbf{Q}\mathbf{r})}
    J_{in}(\mathbf{r}, \mathbf{r'}) \mbox d\mathbf{r}\mbox d\mathbf{r'} \right| \ .
	\label{supp::nominator}
\end{equation}
Here $J_{in}(\mathbf{r}, \mathbf{r'}) = \langle E_{in}^{\mbox*}(\mathbf{r'}) E_{in}(\mathbf{r}) \rangle$ is the MIF function of the incoming field at the sample position, where $E_{in}(\mathbf{r})$  is the complex amplitude of the incident beam.

It is important to note that the contrast function $\zeta_2(\sigma_{\omega})$ in Eq.~\eqref{supp::g2_final_spectr} is defined by the monochromator settings and its value is preserved between the sample and detector positions.
This allows to connect pulse duration of the beam to coherence time of the monochromatized beam incident at the sample.
The functional dependence of the intensity-intensity correlation function as given by Eq.~\eqref{supp::g2_final_spectr} allows to study both the spatial and temporal statistical properties of the XFEL radiation by the HBT interferometry.

\begin{figure}[h]
        \includegraphics[width=10cm]{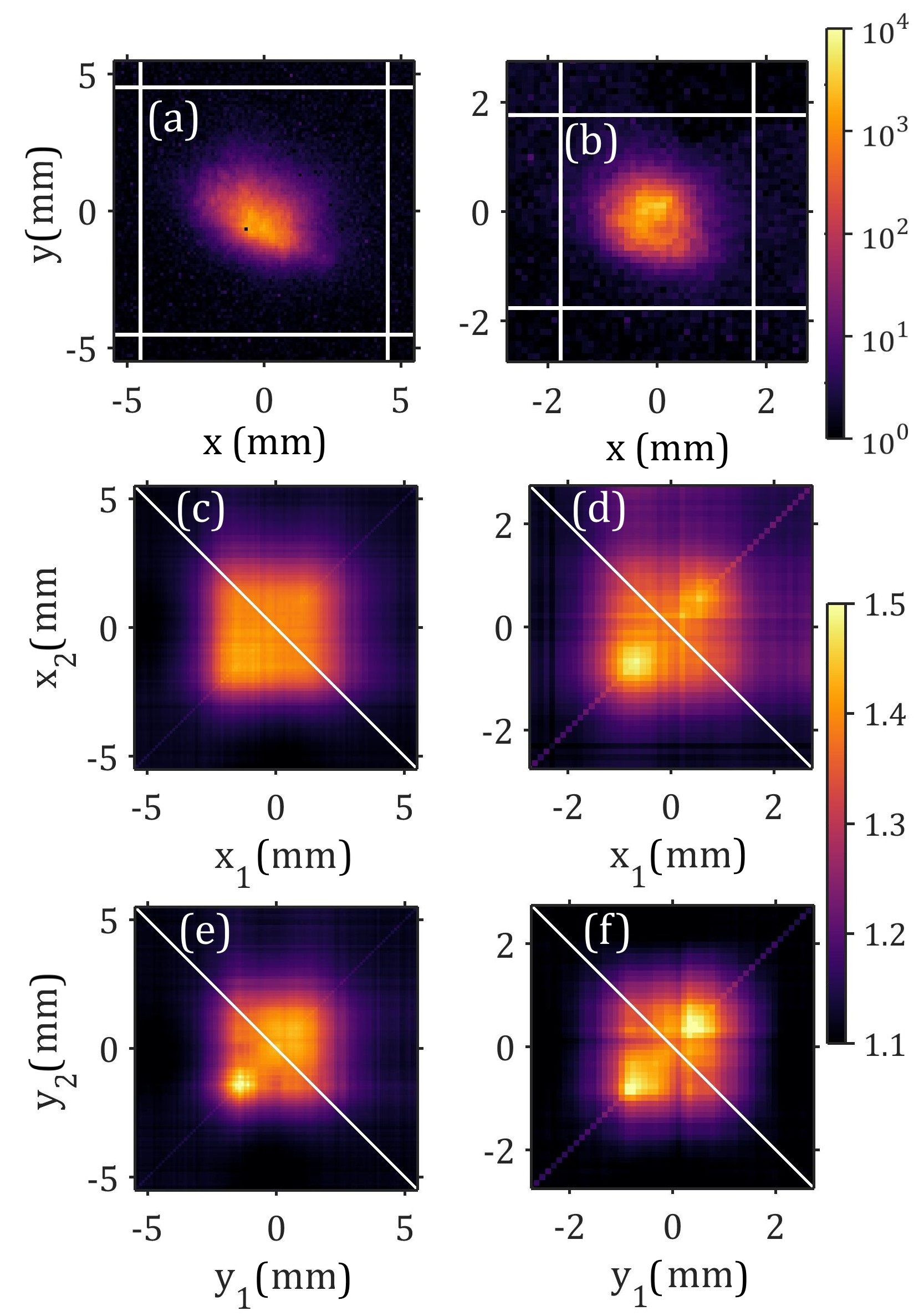}
        \caption{\label{fig::g2}
                (Color online)
                (a-b) Examples of average Bragg peak intensities (shown in log-scale) for sample 1 (a) and sample 2 (b) (peak 4 for both samples).
                (c-f)  Intensity-intensity correlation functions  $g^{(2)}(x_1,x_2)$ (c,d) and $g^{(2)}(y_1,y_2)$ (e,f) evaluated for the same peak 4 and corresponding to sample 1 (c,e) and sample 2 (d,f), respectively.
        }
        \newpage
\end{figure}

For correlation analysis we considered four Bragg peaks not obscured by detector gaps for each crystal (see Fig.~\ref{fig::Scheme} (b-c)).
In order to exclude the influence of the electron energy jitter, only patterns corresponding to pulses with the electron energies close to the mean value were selected (in total about $50,000$, see Appendix for details).
Average intensities of the Bragg peaks marked as 4 in Fig.~\ref{fig::Scheme}(b-c) for both crystals are presented in Fig.~\ref{fig::g2} (a-b).
Projections of the selected Bragg peaks on the horizontal and vertical axes were then correlated and corresponding intensity-intensity correlation functions are presented in (Fig.~\ref{fig::g2} (c-f)).
The comparison of the intensity profiles with the intensity-intensity correlation functions reveals an interesting feature.
While intensity profiles are not smooth and contain several sub-peaks reflecting non-perfect structure of the colloidal crystals, correlation functions are almost flat in a wide central region and then drop down fast to a background level, forming a square type shape (see Appendix for details).
This is different from our earlier measurements at FELs~\cite{Singer2013, Singer2013_Erratum, Gorobtsov2017}, when intensity-intensity correlation functions had been gradually decreasing with the distance between correlated positions.

\begin{figure}[h]
        \includegraphics[width=10cm]{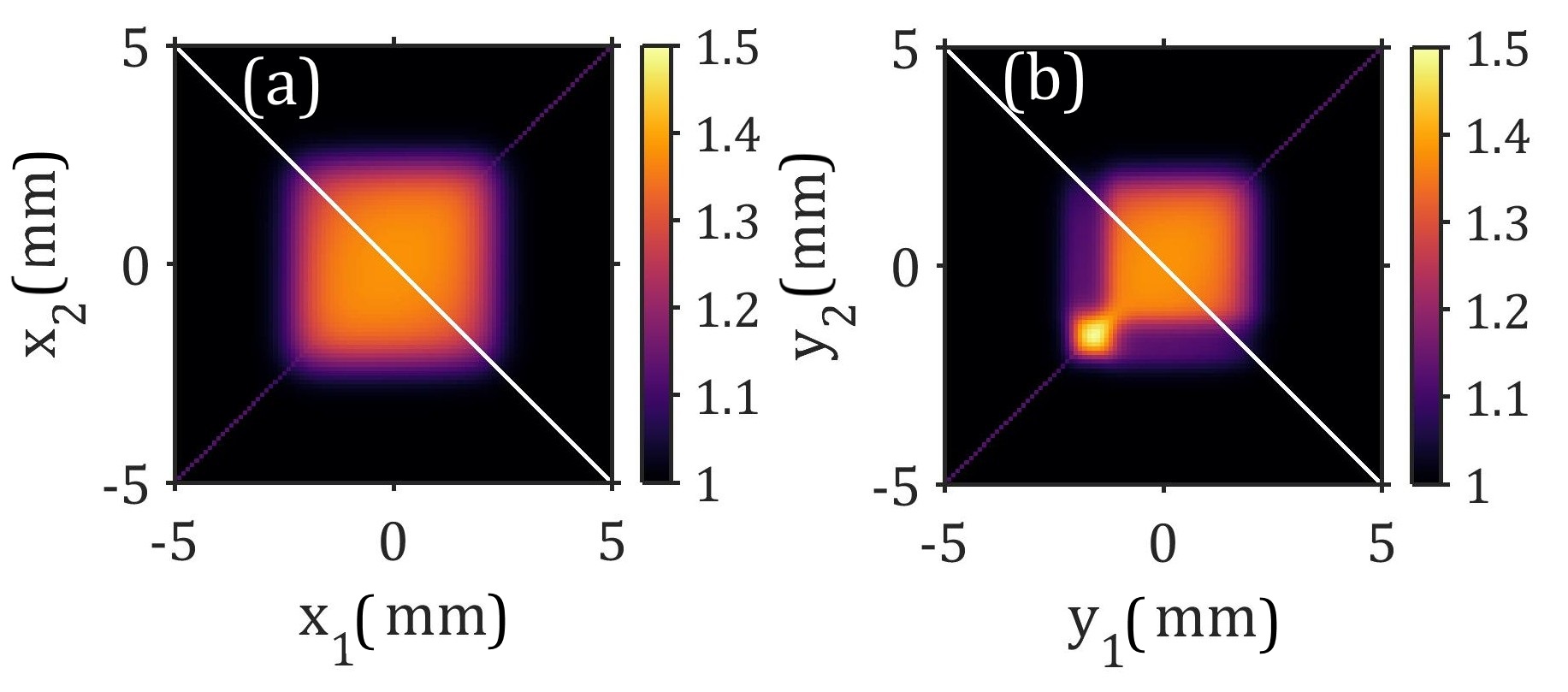}
        \caption{\label{fig::cross}
                (Color online)
                Simulated intensity-intensity correlation functions.
                (a) Single beam with a value of the spatial coherence length $10$ mm which is much larger than the beam size (FWHM) of $1.6$ mm and an additional background noise of $2$\% of the maximum intensity.
                (b) Strong main and a weak secondary beam with the same background noise. Secondary beam has $10$\% of the intensity of the main beam.
        }
        \newpage
\end{figure}

We were able to reproduce this form of intensity-intensity correlation functions in simulations (see Fig.~\ref{fig::cross}).
Two factors contribute to it: coherence length larger than the beam size and additional detector noise.
If a fluctuating background is present on the detector, it limits the field of view of the correlation function to the area where the intensity around the Bragg peak is larger than the detector noise.
If the coherence length of the incident beam is at least a few times larger than the size of the beam, it leads to a relatively flat intensity-intensity correlation function.
We were also able to reproduce an appearance of a small area of higher contrast observed in Fig.~\ref{fig::g2}(e).
It can be simulated using the model of secondary beams introduced in \cite{Gorobtsov2017}.
A weak secondary beam (10\% of the primary beam intensity) introduced in the vertical direction (see for its characteristics Appendix) leads to a similar feature in the intensity-intensity correlation function as observed in the experiment (see Fig.~\ref{fig::cross}(b)).
The fact that the models based on the assumption of a chaotic source describe well the behavior of the intensity-intensity correlation function supports an assumption that LCLS as a SASE XFEL can be considered as a rather chaotic source (compare with \cite{Singer2013}).

Our experimental results also allowed us to determine the degree of spatial coherence of LCLS radiation for hard x-rays.
Performing similar analysis as in Refs.~\cite{Singer2013,Singer2013_Erratum,Gorobtsov2017} we determined that the spatial degree of coherence on average for both samples and different peaks for each direction (horizontal and vertical) was about $0.90 \pm 0.05$ (see Appendix for details). This gave us an estimate of 81\% of global transverse coherence of the full beam, which is in a good agreement with our previous observations \cite{Vartanyants2011, Vartanyants2016, Gorobtsov2017}.

We now explore the temporal properties of the beam.
The contrast $\zeta_2(\sigma_{\omega})$ introduced in Eq.~\eqref{supp::g2_final_spectr} can be determined from the values of the intensity-intensity correlation function along the main diagonal $g^{(2)}(x, x)$ (Fig.~\ref{fig::g2}).
In our experiment it was  approximately $0.40 \pm 0.05$ and did not change significantly for different crystals and Bragg peaks (exact numbers for each crystal and peak can be found in Appendix).
This suggests that the influence of the crystal structure variations on our results is insignificant.
Assuming Gaussian Schell-model pulsed source~\cite{Lajunen2005} (see Appendix for details) the contrast function $\zeta_2(D_{\omega})$ can be expressed as
\begin{equation}
\zeta_2(\sigma_{\omega}) = \frac{1}{\sqrt{1 + 4(T_{rms} \sigma_{\omega})^2}} \ ,
\label{contrast_function}
\end{equation}
where $T_{rms}$ is an effective pulse duration (r.m.s.) before the monochromator and $\sigma_{\omega}$ is the r.m.s. value of the monochromator bandwidth.
It is important to point out that the effective pulse duration is extracted only from the part of the beam passing the monochromator.
As such, it can be significantly shorter than that of non- monochromatized beam (see below).
In derivation of equation~\eqref{contrast_function} it was assumed that the spectral width of the incoming radiation is much broader than the monochromator bandwidth and spectral coherence width.
These conditions are well satisfied for the LCLS x-ray beam parameters and monocromator used in the experiment.
Inversion of equation~\eqref{contrast_function} gives for the FWHM of the pulse duration (see Appendix for details)
%
\begin{equation}
T = \frac{2.355}{2\sigma_{\omega}}\sqrt{\frac{1}{\left[\zeta_2(\sigma_{\omega})\right]^2}-1} \ .
\label{duration_zeta}
\end{equation}
Using the measured value of the contrast function $\zeta_2(\sigma_{\omega})$, we can estimate that for our experiment the pulse duration lies in the range of $11-12$ fs.
These values were significantly shorter than initially expected (about $41$ fs) from the electron bunch measurements.

To verify our findings we determined pulse duration by a different approach based on the mode analysis of the radiation as suggested in  Ref.~\cite{Saldin1998}.
According to this approach an average number of modes of radiation $M$ is inversely proportional to the normalized dispersion of the energy distribution, that in our case coincide with the contrast function defined in Eq.~\eqref{supp::g2_final_spectr} $M=1/\left[\zeta_2(\sigma_{\omega})\right]$.
We determined the number of modes by fitting intensity distribution at one of the Bragg peaks by Gamma distribution~\cite{Saldin1998} (see Appendix for details).
As a result, the number of longitudinal modes was $M \approx 2.3\pm0.1$ and reproducible between different runs.
Substituting this number in Eq.~\eqref{duration_zeta} gives for the pulse duration $11.5 \pm 0.5$ fs in an excellent agreement to previously determined values from the HBT interferometry.
Similar inconsistency factor of about three between the expected and observed pulse duration has been observed earlier in another LCLS experiment~\cite{Gutt2012}.

To explain the difference between thus obtained values of the pulse duration with the results of the electron bunch measurements several factors should be taken into account.
An estimate of the pulse duration from the electron bunch measurements is mainly based on the longitudinal size of the electron beam as an FEL lasing medium.
The electron beam size generally limits the maximum emitted hard x-ray pulse duration.
The FEL gain is very sensitive to various electron beam properties, such as beam emittance, electron current and energy spread, and orbit alignment inside an undulator.
These properties vary along the electron beam, which may result in a relatively short core, providing significantly better gain, compared to the rest of the beam.
Another possible explanation may be the filtering of the bunch with the strong chirp by the high-resolution monochromator~\cite{Krinsky2003}.
It was proven experimentally with cross-correlation measurements~\cite{Ding2012}, that 150 pC 50 fs long electron beam may radiate only 14 fs long 8.5~keV beam, which is comparable with our observations.

The coherence time $\tau_c$ can be estimated from the bandwidth of the monochromator according to Ref.~\cite{Saldin1998} as
$\tau_c = \sqrt{\pi}/\sigma_{\omega}$.
%
%
The obtained value is about $7.5$ fs, which is only slightly shorter than the pulse duration.
Therefore, x-ray pulses were effectively longitudinally coherent during the experiment.

In summary, we have performed HBT interferometry at the Bragg peaks originating from the colloidal crystals measured at LCLS.
This technique allowed us to extract information about spatial coherence and temporal properties of the incident beam directly from the diffraction patterns without additional equipment or specially dedicated measurements.
We have determined a high degree of spatial coherence of the full XFEL beam that was about $81\%$ which concord with our previous measurements.
We have also observed a coherence length much larger than the beam size.
We have obtained pulse durations of $11-12$ fs, which are significantly shorter than expected in the operation regime of the LCLS used in our experiment.
We also estimated coherence time for high-resolution monochromator used in our experiment and obtained the value of $7.5$ fs that is just slightly below the pulse duration.
That means that LCLS pulses in our experiment were not only spatially but also temporally coherent close to Fourier limited pulses.

Our approach is quite general and is not limited to the analysis of the diffraction patterns originating from colloidal crystals.
Any other crystalline sample can be used provided Bragg peaks are sufficiently broad to allow HBT measurement.
This can be accommodated, for example, by the larger sample detector distance, or implementing a set of CRLs in the beam diffracted from the sample.

Our measurements have demonstrated high degree of spatial coherence of the FEL radiation that could potentially lead to completely new avenue in the field of quantum optics.
Such quantum optics experiments as exploration of non-classical states of light \cite{Breitenbach1997},
super-resolution experiments \cite{Oppel2012}, quantum imaging experiments \cite{Schneider2017} or ghost imaging experiments \cite{Pelliccia2016,Yu2016} may become possible at the hard XFEL sources in the near future.
Finally, we foresee that HBT interferometry will become an important diagnostics and analytic tool at the XFEL sources.


Portions of this research were carried out on the XPP Instrument at the LCLS at the SLAC National Accelerator Laboratory.
LCLS is an Office of Science User Facility operated for the U.S. Department of Energy Office of Science by Stanford University.
Use of the Linac Coherent Light Source (LCLS), SLAC National Accelerator Laboratory, is supported by the U.S. Department of Energy, Office of Science, Office of Basic Energy Sciences under Contract No. DE-AC02-76SF00515.
We thank Edgar Weckert, Evgeny Saldin, Dina Sheyfer and Gerhard Gr{\"u}bel for useful discussions.
This work was partially supported by the Virtual Institute VH-VI-403 of the Helmholtz Association.

\bibliography{references}

\eject

\appendix

\section{Appendix I. Intensity-intensity correlation functions of a radiation field scattered from a crystal}

\begin{figure}[h]
	\includegraphics[width=10cm]{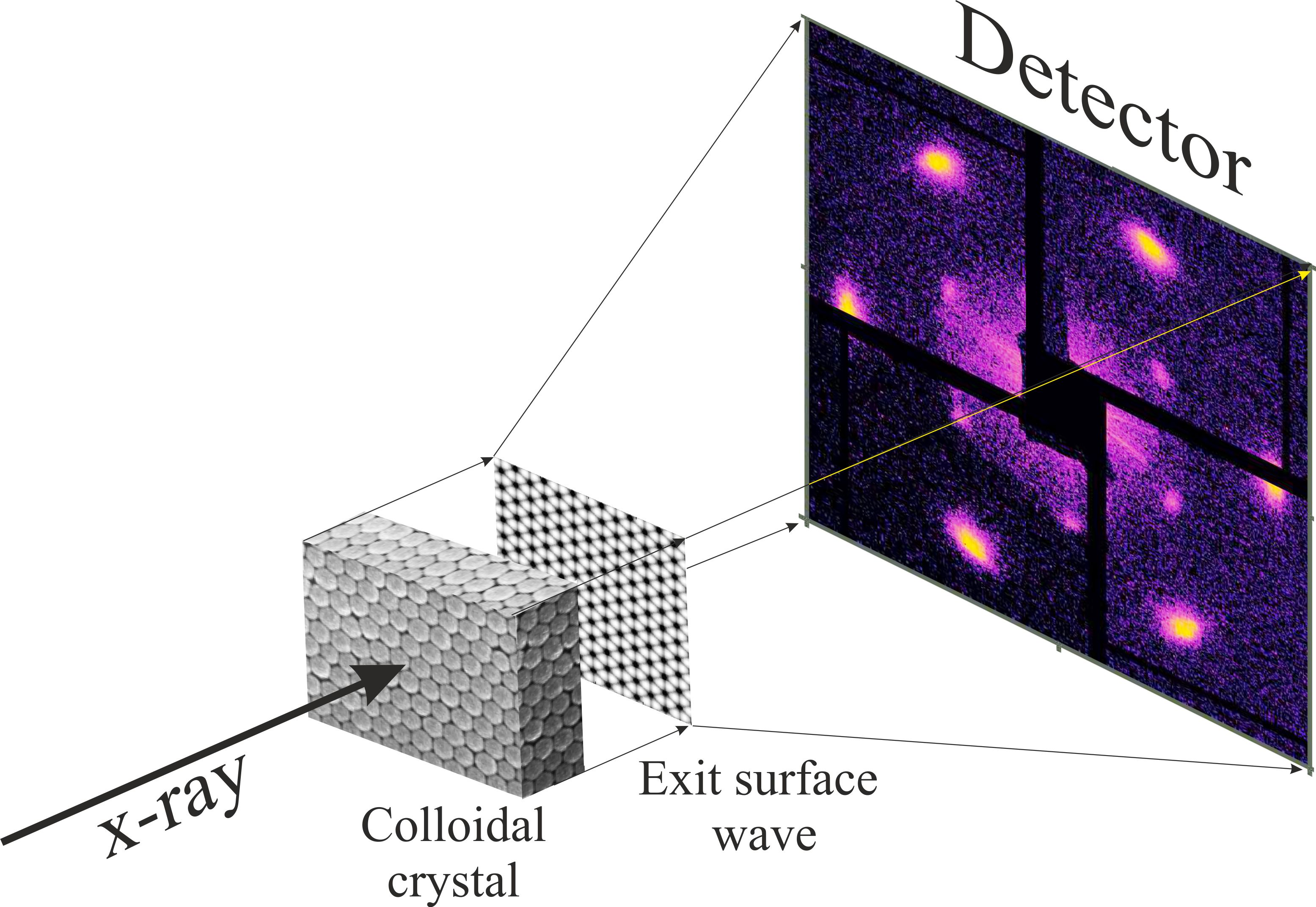}
	\caption{\label{fig::Diffraction_scheme}
		Scheme of diffraction experiment on a colloidal crystal performed at LCLS.
		An x-ray beam from the LCLS is incoming on a thin colloidal crystal film, just behind a film an exit surface wave is formed that is propagating in free space towards detector position.
	}
	\newpage
\end{figure}

We will consider a quasi-monochromatic x-ray beam $E_{in}(\mathbf{s})$ incident on a colloidal crystal in a shape of a thin slab of material (see Fig.\ref{fig::Diffraction_scheme}).

An exit surface wave from such a crystal can be written as
\begin{equation}
	E_{ESW}(\mathbf{s}) = O(\mathbf{s})E_{in}(\mathbf{s}) \ ,
	\label{supp::object_function}
\end{equation}
where $O(\mathbf{s})$ is the so-called object function and $\mathbf{s}$ is the two-dimensional (2D) vector in transverse direction to the incoming beam at the position of the sample.
For a thin slab of material an object function can be expressed through refractive index $n(\mathbf{s},z)$ as~\cite{Als-Nielsen}
\begin{equation}
O(\mathbf{s}) = e^{i\varphi(\mathbf{s})} \ ,
\label{supp::object_refr}
\end{equation}
where $\varphi(\mathbf{s}) = k\int_{0}^{d(\mathbf{s})}(n(\mathbf{s},z)-1)\mbox{d}z$ is the phase difference due to refraction.
Here $d(\mathbf{s})$ is the crystal thickness at the position $\mathbf{s}$, $k=2\pi/\lambda$ is the wave number and $\lambda$ is the wavelength.
At x-ray wavelength refractive index can be expressed as~\cite{Als-Nielsen} $n(\mathbf{s},z) = 1-\delta(\mathbf{s},z)+i\beta(\mathbf{s},z)$, where $\delta(\mathbf{s},z)$ is the real part of refractive index responsible for refraction and $\beta(\mathbf{s},z)$ is the imaginary part responsible for absorption.
Neglecting absorption and taking into account known relation between the real part of refractive index and electron density of the crystal~\cite{Als-Nielsen} $\delta(\mathbf{s},z) = \lambda r_e\rho(\mathbf{s},z)/k$, where $r_e$ is the classical electron radius, we obtain for the phase in the object function in Eq.~\eqref{supp::object_refr}
\begin{equation}
\varphi(\mathbf{s}) = - \lambda r_e\int_{0}^{d(\mathbf{s})}\rho(\mathbf{s},z)\mbox{d}z.
\label{supp::phase_diff}
\end{equation}
Taking into account that projection of a crystalline electron density is a periodic function we obtain that the object function in Eq.~\eqref{supp::object_refr} is also 2D periodic function.

To determine distribution of the wavefield at the detector position we will propagate the exit surface wave to that position by performing convolution with the free space propagator $P_L(\mathbf{r})$
\begin{equation}
E_{d}(\mathbf{r})   = \int E_{ESW}(\mathbf{s})P_L(\mathbf{r}-\mathbf{s})\mbox{d}\mathbf{s}
                    = \int O(\mathbf{s}) P_L(\mathbf{r}-\mathbf{s}) E_{in}(\mathbf{s}) \mbox{d}\mathbf{s} \ ,
\label{supp::prop_wave}
\end{equation}
where $\mathbf{r}$ is the 2D coordinate at the detector position and $L$ is the sample-detector distance.
Propagator in Eq.~\eqref{supp::prop_wave} has a known form
\begin{equation}
P_L(\mathbf{r}-\mathbf{s}) = \frac{1}{i \lambda L}\exp \left[ik\frac{(\mathbf{r}-\mathbf{s})^2}{2L}\right] \ .
\label{supp::propagator}
\end{equation}

Taking now into account that the object function is a 2D periodic function it can be expanded into Fourier series as
\begin{equation}
O(\mathbf{s}) = \sum_{\mathbf{h}} O_{\mathbf{h}} e^{i\mathbf{h} \cdot \mathbf{s}} \ ,
\label{supp::h_sum}
\end{equation}
where $\mathbf{h}$ is the 2D reciprocal space vector and
$O_{\mathbf{h}} = 1/V \int O(\mathbf{s}) e^{-i \mathbf{h} \cdot \mathbf{s}}\mbox{d}\mathbf{s}$
are the Fourier coefficients of the expansion.
Substituting now this expansion in Eq.~\eqref{supp::prop_wave} and considering scattering in the vicinity of a selected Bragg peak $\mathbf{h}$ we obtain
\begin{equation}
E_h(\mathbf{s}) = O_{\mathbf{h}} \int e^{i\mathbf{h} \cdot \mathbf{s}} P_L(\mathbf{r}-\mathbf{s}) E_{in}(\mathbf{s}) \mbox{d}\mathbf{s} \ .
\label{supp::one_peak}
\end{equation}

Using the far-field ($D^2/(\lambda L) \gg 1$, where $D$ is the size of the beam at the sample position) expression of the propagator $P_L(\mathbf{r}-\mathbf{s}) \simeq exp(-i\mathbf{q_r} \cdot \mathbf{s})$, where $\mathbf{q_r} = k \mathbf{r}/L$ we obtain from Eq.~\eqref{supp::one_peak}
\begin{equation}
E_h^{FF}(\mathbf{Q}) = O_{\mathbf{h}} \int e^{-i \mathbf{Q} \cdot \mathbf{s}}  E_{in}(\mathbf{s}) \mbox{d}\mathbf{s} \ ,
\label{supp::one_peak_far-field}
\end{equation}
where $\mathbf{Q} = \mathbf{q_r} - \mathbf{h}$ is the momentum transfer vector calculated from the reciprocal space vector $\mathbf{h}$.
For the intensity of the scattered field in the far-field we have
\begin{equation}
I_h^{FF}(\mathbf{Q})   = \left|E_h(\mathbf{Q}) \right|^2
                    = \left|O_{\mathbf{h}}\right|^2 \int \int e^{-i \mathbf{Q} \cdot (\mathbf{s}-\mathbf{s'})} E_{in}^*(\mathbf{s'}) E_{in}(\mathbf{s}) \mbox{d}\mathbf{s} \mbox{d}\mathbf{s'} \ .
\label{supp::intens_far-field}
\end{equation}
In Fresnel (near-field) regime we can not use expansion expression for the propagator and we have for the scattered amplitude
\begin{equation}
E_h^{NF}(\mathbf{Q}) = O_{\mathbf{h}} e^{i\phi_u} \int e^{-i \mathbf{Q} \cdot \mathbf{s}}  \widetilde{E}_{in}(\mathbf{s}) \mbox{d}\mathbf{s} \ ,
\label{supp::one_peak_near-field}
\end{equation}
where we introduced the phase $\phi_r=k\mathbf{r}^2/(2L)$ and defined a new amplitude
\begin{equation}
\widetilde{E}_{in}(\mathbf{s}) = e^{i (k/2L)\mathbf{s}^2} E_{in}(\mathbf{s}) \ .
\label{supp::inc_ampl_near-field}
\end{equation}
For intensity in the near-field we have
\begin{equation}
I_h^{NF}(\mathbf{Q})   = \left|E_h^{NF}(\mathbf{Q}) \right|^2
                    = \left|O_{\mathbf{h}}\right|^2 \int \int e^{-i \mathbf{Q} \cdot (\mathbf{s}-\mathbf{s'})} \widetilde{E}_{in}^*(\mathbf{s'}) \widetilde{E}_{in}(\mathbf{s}) \mbox{d}\mathbf{s} \mbox{d}\mathbf{s'} \ .
\label{supp::intens_near-field}
\end{equation}
As we can see expressions for the scattered intensities around selected Bragg peak coincide in the far-ield and near-field conditions with the change of the incoming wavefield expression to one given in Eq.~\eqref{supp::inc_ampl_near-field}.
As soon as the difference between two cases is in the constant phase factor it would not influence statistical characteristics of the scattered field. In the following we will use far-field expression~\eqref{supp::intens_far-field} keeping in mind that Fresnel conditions can be matched by the substitution given in Eq.~\eqref{supp::inc_ampl_near-field}.

We will now evaluate intensity-intensity correlation function at the detector position in the vicinity of the selected Bragg reflection $\mathbf{h}$
\begin{equation}
g^{(2)}(\mathbf{Q}, \mathbf{Q'})
    = \frac{\langle I(\mathbf{Q}) I(\mathbf{Q'}) \rangle}{\langle I(\mathbf{Q})\rangle \langle I(\mathbf{Q'}) \rangle} \ ,
\label{supp::2_order_normalized}
\end{equation}
where momentum transfer vectors $\mathbf{Q}$ and $\mathbf{Q'}$ are centered at reflection $\mathbf{h}$ and related to the spatial coordinates at the detector position by $\mathbf{Q} = k \mathbf{u}/L$ and $\mathbf{Q'} = k \mathbf{u'}/L$.
Averaging here is denoted by the brackets $<...>$ and is performed over many realizations of the field.
Substituting here expression for the intensity~\eqref{supp::intens_far-field} we have for the nominator

\begin{widetext}
	\begin{multline}
	\langle I(\mathbf{Q}) I(\mathbf{Q'}) \rangle =
    |O_{\mathbf{h}}|^4
	\iiiint  e^{-i\mathbf{Q} \cdot (\mathbf{s}-\mathbf{s'}) - \mathbf{Q'} \cdot (\mathbf{s''}-\mathbf{s'''})}
    \langle E_{in}^{\mbox*}(\mathbf{s'}) E_{in}(\mathbf{s}) E_{in}^{\mbox*}(\mathbf{s'''}) E_{in}(\mathbf{s''})\rangle
    \mbox d\mathbf{s}\mbox d\mathbf{s'}\mbox d\mathbf{s''}\mbox d\mathbf{s'''} \ .
	\label{supp::nominator}
	\end{multline}
\end{widetext}
Assuming that the incoming radiation obeys Gaussian statistics we can use Gaussian moment theorem
\begin{widetext}
	\begin{multline}
	\langle E_{in}^{\mbox*}(\mathbf{s'}) E_{in}(\mathbf{s}) E_{in}^{\mbox*}(\mathbf{s'''}) E_{in}(\mathbf{s''})\rangle
    = \langle E_{in}^{\mbox*}(\mathbf{s'}) E_{in}(\mathbf{s})\rangle \langle E_{in}^{\mbox*}(\mathbf{s'''}) E_{in}(\mathbf{s''})\rangle + \\
	+ \langle E_{in}^{\mbox*}(\mathbf{s'}) E_{in}(\mathbf{s''})\rangle \langle E_{in}^{\mbox*}(\mathbf{s'''}) E_{in}(\mathbf{s})\rangle \ .
	\label{supp::gauss_theorem}
	\end{multline}
\end{widetext}
Substituting now this expression in Eq.~\eqref{supp::nominator} we obtain for the nominator
\begin{equation}
	\langle I(\mathbf{Q}) I(\mathbf{Q'}) \rangle
    = \langle I(\mathbf{Q}) \rangle \langle I(\mathbf{Q'}) \rangle +
    \left| J(\mathbf{Q}, \mathbf{Q'}) \right|^2 \ ,
	\label{supp::nominator}
\end{equation}
where $|J(\mathbf{Q}, \mathbf{Q'})|$ is the absolute value of the mutual intensity function (MIF) defined at the detector position and related to the MIF of the incoming field
$J_{in}(\mathbf{s}, \mathbf{s'}) = \langle E_{in}^{\mbox*}(\mathbf{s'}) E_{in}(\mathbf{s}) \rangle$
by the following relation
\begin{equation}
	\left| J(\mathbf{Q}, \mathbf{Q'}) \right|^2  = \left|\iint e^{-i(\mathbf{Q'}\mathbf{s'}-\mathbf{Q}\mathbf{s})}
    J_{in}(\mathbf{s}, \mathbf{s'}) \mbox d\mathbf{s}\mbox d\mathbf{s'} \right|^2 \ .
	\label{supp::nominator}
\end{equation}
Finally, we have for the normalized intensity-intensity correlation function~\eqref{supp::2_order_normalized}
\begin{equation}
g^{(2)}(\mathbf{Q}, \mathbf{Q'})
    = \frac{\langle I(\mathbf{Q}) I(\mathbf{Q'}) \rangle}{\langle I(\mathbf{Q})\rangle \langle I(\mathbf{Q'}) \rangle}
    = 1+\left| \mu(\mathbf{Q}, \mathbf{Q'})\right|^2 \ ,
\label{supp::g2_final}
\end{equation}
where
\begin{equation}
\mu(\mathbf{Q}, \mathbf{Q'}) = \frac{J(\mathbf{Q}, \mathbf{Q'})}{\sqrt{\langle I(\mathbf{Q}) \rangle}\sqrt{\langle I(\mathbf{Q'}) \rangle}}
\label{supp::SDC}
\end{equation}
is the normalized spectral degree of coherence.

Taking now into account that we have a finite bandwidth of radiation incoming from the monochromator we have for the intensity-intensity correlation function
\begin{equation}
g^{(2)}(\mathbf{Q}, \mathbf{Q'})= 1 + \zeta_2(\sigma_{\omega})\left| \mu(\mathbf{Q}, \mathbf{Q'})\right|^2 \ ,
\label{supp::g2_final_spectr}
\end{equation}
where $\zeta_2(\sigma_{\omega})$ is the contrast function which strongly depends on the radiation bandwidth $\sigma_{\omega}$.
We will now evaluate this function in  the next section.

\section{Appendix II. Determination of the pulse duration from the intensity interferometry}

In the HBT interferometry the contrast function $\zeta_2(\sigma_{\omega})$ for a cross-spectral pure chaotic radiation can be defined as \cite{Singer2013, Singer2013_Erratum, Gorobtsov2017}
\begin{equation}
\zeta_2(\sigma_{\omega}) = \frac{\iint\limits_{-\infty}^{\infty} |T(\omega_1)|^2|T(\omega_2)|^2|W(\omega_1,\omega_2)|^2 \mbox d\omega_1 \mbox d\omega_2}
{\left[\int\limits_{-\infty}^{\infty}|T(\omega)|^2S(\omega)\mbox d\omega \right]^2}\ ,
\label{coherence::zeta_2}
\end{equation}
where $|T(\omega)|^2$ is the monochromator transmission function, $W(\omega_1,\omega_2)$ is the cross spectral density function in the spectral domain, and $S(\omega) = W(\omega,\omega)$ is the spectral density function.

We will assume in the following that monochromator transmission function is described by a Gaussian function with the r.m.s. width $\sigma_{\omega}$
\begin{equation}
|T(\omega)|^2 = \exp\left[ \frac{\omega^2}{2\sigma_{\omega}^2}\right]
\label{coherence::transmission}
\end{equation}
and pulsed x-ray radiation incoming on the monochromator can be approximated by a Gaussian Schell-model beam giving for the cross spectral density function~\cite{Lajunen2005}
%
%
%
%
\begin{equation}
W(\omega_1,\omega_2)  =
    W_{0} \exp\left[-\frac{(\omega_1-\omega_0)^2+(\omega_2-\omega_0)^2}{4\Omega^2}-\frac{(\omega_1-\omega_2)^2}{2\Omega_c^2}\right] \ ,
\label{coherence::GSM_JM}
\end{equation}
where $W_{0}$ is the normalization constant.
Here $\omega_0$ is the central pulse frequency, $\Omega$ is the spectral width, and $\Omega_c$ is the spectral coherence width.
It can be shown~\cite{Lajunen2005} that these parameters can be related to the r.m.s. values of the pulse duration $T_{rms}$ and coherence time $T_c$ of the pulse before monochromator as~\cite{Lajunen2005}
\begin{equation}
\Omega^2 = \frac{1}{T_c^2} + \frac{1}{4T_{rms}^2} \ ; \  \Omega_c = \frac{T_c}{T_{rms}}\Omega \ .
\label{coherence::band_time}
\end{equation}

Now substituting Eqs.~(\ref{coherence::transmission} - \ref{coherence::band_time}) into the expression for the contrast function Eq.~\eqref{coherence::zeta_2} and performing integration we obtain
\begin{equation}
\zeta_2(\sigma_{\omega}) = \frac{2C}{\sqrt{4A^2-B^2}} \ ,
\label{coherence::zeta_2_general}
\end{equation}
where
\begin{equation}
A = \frac{1}{2\sigma_{\omega}^2} + \frac{1}{2\Omega^2} + \frac{1}{\Omega_c^2}  \ ;
B = \frac{2}{\Omega_c^2} \ ;
C = \frac{1}{2\sigma_{\omega}^2} + \frac{1}{2\Omega^2} \ .
\label{coherence::parameters}
\end{equation}
This is the general expression for the contrast function for arbitrary values of all frequencies introduced in this expression.
Now taking into account that in the conditions of our experiment at LCLS the monochromator bandwidth $\sigma_{\omega}$ and spectral coherence width $\Omega_c$ were much narrower than the spectral width $\Omega$ ($\sigma_{\omega}, \Omega_c \ll \Omega$) we obtain for parameters~\eqref{coherence::parameters} the following approximate expression
\begin{equation}
A \simeq \frac{1}{2\sigma_{\omega}^2}  + \frac{1}{\Omega_c^2}  \ ;
B = \frac{2}{\Omega_c^2} \ ;
C \simeq \frac{1}{2\sigma_{\omega}^2}  \ .
\label{coherence::parameters_approx}
\end{equation}
Substituting these values in expression~\eqref{coherence::zeta_2_general} we obtain for the contrast function
\begin{equation}
\zeta_2(\omega) = \frac{\Omega_c}{\sqrt{\Omega_c^2+4\sigma_{\omega}^2}}
                = \frac{1}{\sqrt{1+4\left(\sigma_{\omega}/\Omega_c\right)^2}} \ .
\label{coherence::zeta_2_approx}
\end{equation}
Taking now into account that in the conditions of our experiment at LCLS coherence time of radiation before the monochromator was much shorter than the pulse duration ($T_c \ll T_{rms}$) we obtain from Eqs.~\eqref{coherence::band_time} for the pulse duration
\begin{equation}
T_{rms} \simeq \frac{1}{\Omega_c} \ .
\label{coherence::pulse duration}
\end{equation}
Substituting this expression in Eq.~\eqref{coherence::zeta_2_approx} we obtain for the contrast function the following relation
\begin{equation}
\zeta_2(\omega) = \frac{\Omega_c}{\sqrt{\Omega_c^2+4\sigma_{\omega}^2}}
                = \frac{1}{\sqrt{1+4\left(T_{rms} \sigma_{\omega}\right)^2}} \ .
\label{coherence::zeta_2_approx1}
\end{equation}
that was used in the main text of the manuscript for the analysis.
In two limiting cases $T_{rms} \sigma_{\omega} \ll 1$ and $T_{rms} \sigma_{\omega} \gg 1$ we obtain from equation~\eqref{coherence::zeta_2_approx1} for the contrast function: $\zeta_2(\sigma_{\omega}) \simeq 1-2 (T_{rms} \sigma_{\omega})^2$ in the first case and
$\zeta_2(\sigma_{\omega}) \simeq 1/\left[2(T_{rms} \sigma_{\omega})\right]$ in the second.
The first case corresponds to nearly Fourier limited radiation and the second one to rather incoherent (in time-domain) radiation.

Expression~\eqref{coherence::zeta_2_approx1} can be inverted to determine pulse duration of the x-ray radiation before the monochromator.
For the FWHM of the pulse duration we finally have
\begin{equation}
T =  2.355 T_{rms} = \frac{2.355}{2\sigma_{\omega}}\sqrt{\frac{1}{\left[\zeta_2(\sigma_{\omega})\right]^2}-1} \ .
\label{coherence::pulse_duration}
\end{equation}

\section{Appendix III. Additional experimental results}

\begin{figure}[h]
	\includegraphics[width=10cm]{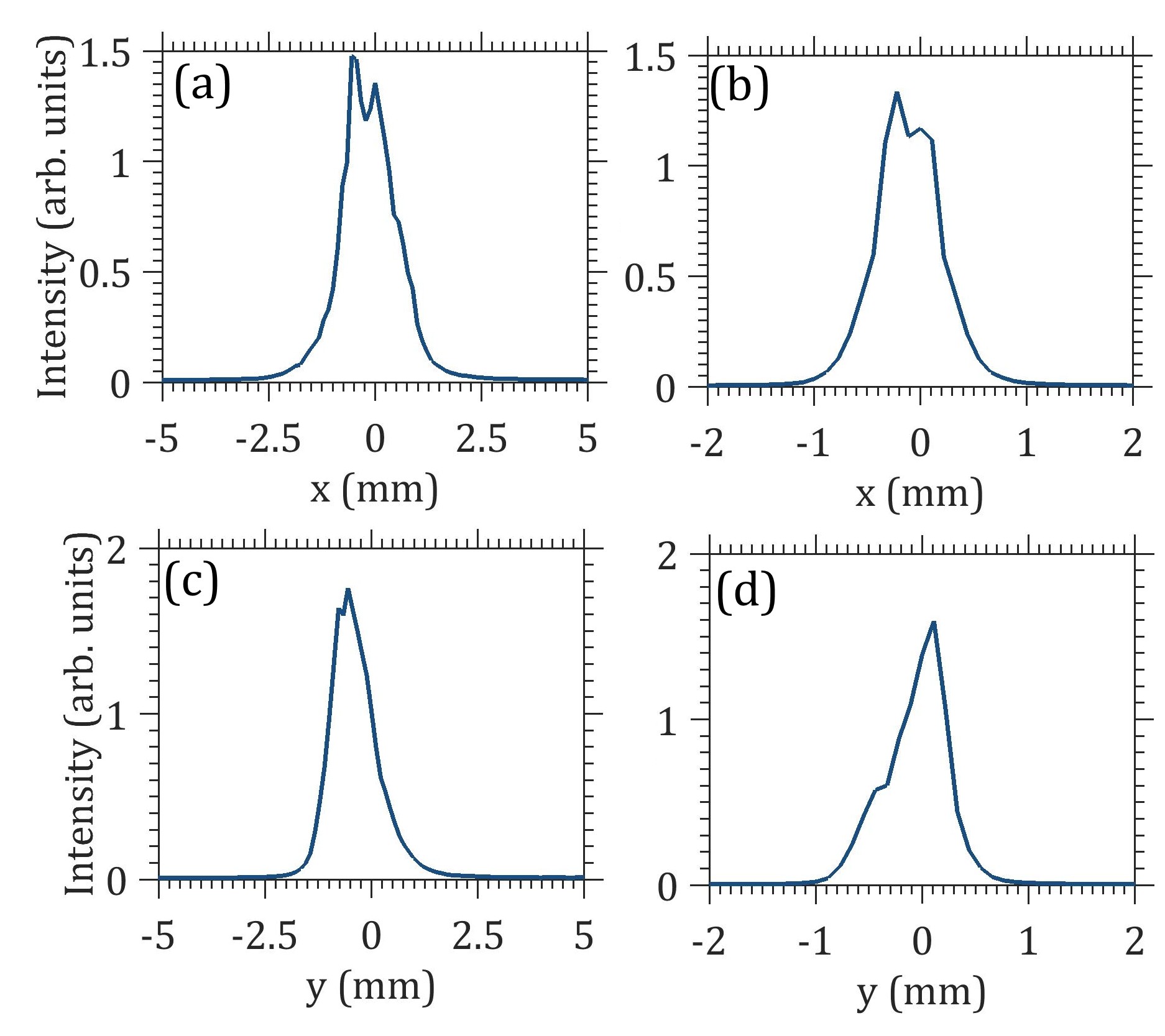}
	\caption{\label{fig::int}
		Projections of the averaged intensity.
		(a, c) Horizontal and vertical projections for the sample 1.
		(b, d) Horizontal and vertical projections for the sample 2.
	}
	\newpage
\end{figure}

\begin{figure}[h]
	\includegraphics[width=10cm]{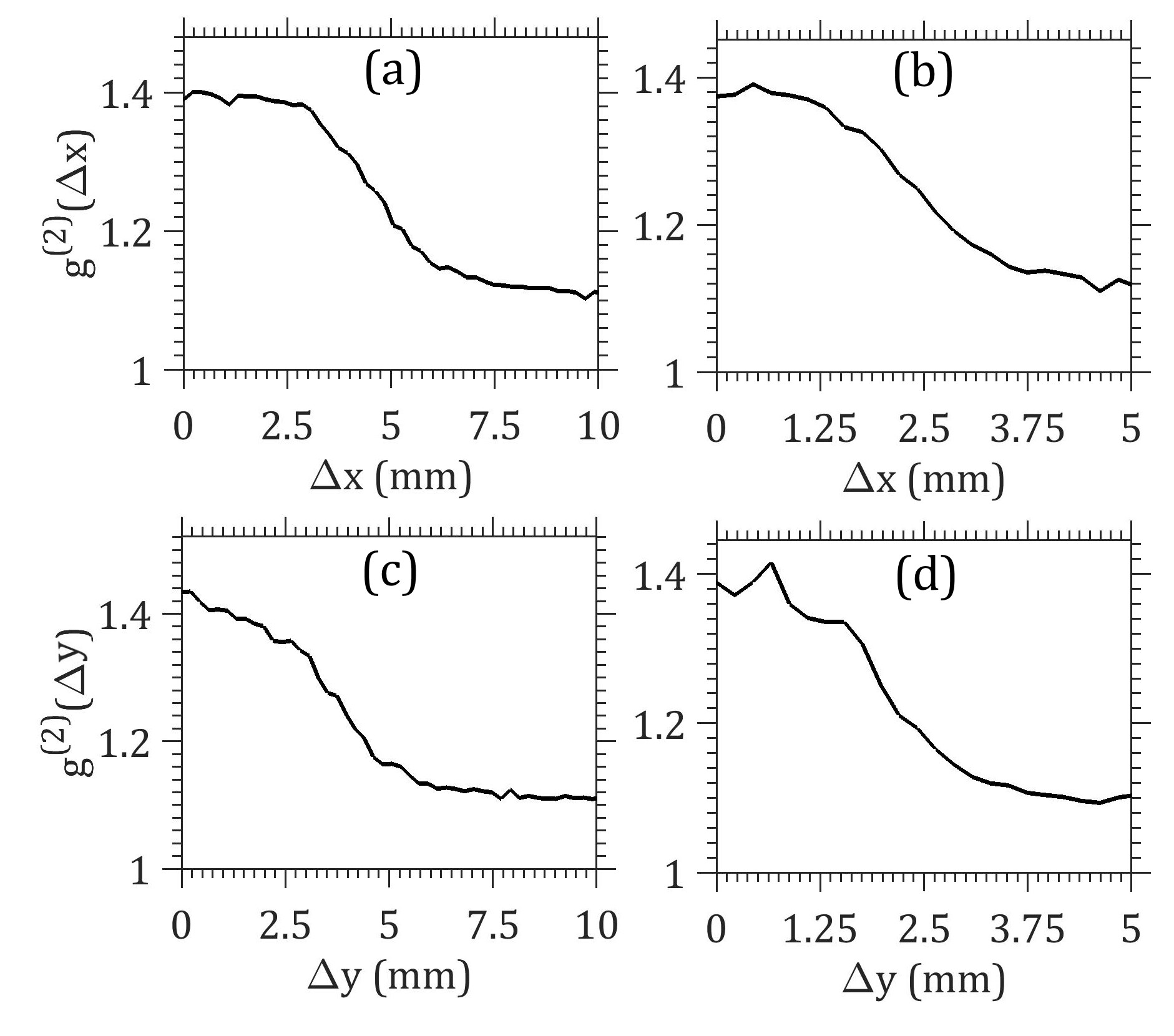}
	\caption{\label{fig::cohcross}
		Intensity-intensity correlation functions  $g^{(2)}(\Delta x)$ and $g^{(2)}(\Delta y)$ taken along the diagonal shown in Fig. 3 (c, f) of the main text as a white line for sample 1 (a, c) and sample 2 (b, d).
	}
	\newpage
\end{figure}

Here we present additional figures demonstrating our experimental results.
Projections of the averaged intensity on the both horizontal and vertical axes (shown in Fig.~\ref{fig::int}) reveal presence of the small subpeaks due to the defect structure of the colloidal crystal.
Cross sections of the intensity-intensity correlation function along the white line in Fig.~3 in the main text (see Fig.~\ref{fig::cohcross}) demonstrate a relatively flat region in the center and a steep slope after that.

\section{Appendix IV. Simulation of the intensity-intensity correlation functions}

The model used for simulations in this work was first introduced in Ref.~\cite{Gorobtsov2017}.
In this model, the X-ray beam is assumed to consist of several statistically independent Gaussian Schell-model beams with the total complex field amplitude
\begin{equation}
E_{\Sigma}(\mathbf{r},\omega)=
\sum_{i=1}^{N} E_i(\mathbf{r},\omega)\ ,
\label{multiple::wave_superposition}
\end{equation}
where $E_i(\mathbf{r},\omega)$ is a complex amplitude of a single beam.
Since all beams are statistically independent, the total spectral cross-correlation function $W^{(2)}_{\Sigma}(\mathbf{r_1},\omega_1,\mathbf{r_2},\omega_2)$ and spectral density $S_{\Sigma}(\mathbf{r},\omega)$ can be expressed as
\begin{eqnarray}
W^{(2)}_{\Sigma}(\mathbf{r_1},\omega_1,\mathbf{r_2},\omega_2)& = & \sum_{i=1}^{N}J_i(\mathbf{r_1}, \mathbf{r_2})W_i(\omega_1, \omega_2)\ , \\
S_{\Sigma}(\mathbf{r},\omega)& = & \sum_{i=1}^{N}I_i(\mathbf{r})S_i(\omega)\ .
\label{multiple::spectral_cross_super}
\end{eqnarray}
Intensity-intensity cross-correlation function is than calculated as obtained in Ref.~\cite{Gorobtsov2017}
\begin{widetext}
	\begin{multline}
	g_{\Sigma}^{(2)}(\mathbf{r_1},\mathbf{r_2})
	=1+\\
	+\frac{\sum_{i,j=1}^{N}J_{i}(\mathbf{r_1},\mathbf{r_2})J^*_{j}(\mathbf{r_1},\mathbf{r_2})\iint\limits_{-\infty}^{\infty}|T(\omega_1)|^2|T(\omega_2)|^2  W_{i}(\omega_1, \omega_2)W^*_{j}(\omega_1, \omega_2)\mbox d\omega_1 \mbox d\omega_2}
	{\sum_{k, l = 1}^{N}I_{k}(\mathbf{r_1})I_{l}(\mathbf{r_2})\int\limits_{-\infty}^{\infty}|T(\omega_1)|^2 S_{k}(\omega_1)\mbox d\omega_1\int\limits_{-\infty}^{\infty}|T(\omega_2)|^2 S_{l}(\omega_2) \mbox d\omega_2}\ .
	\label{multiple::general_g}
	\end{multline}
\end{widetext}

The model for simulating the fluctuating detector background was also introduced in Ref.~\cite{Gorobtsov2017}.
The total intensity can be represented as
\begin{equation}
I(x) = I_0(x) +I_B(x) \ ,
\label{app3::int_and_backg}
\end{equation}
where $I_0(x)$ is the intensity of the beam and $I_B(x)$ is the background intensity.
The background signal is assumed to be statistically independent from the beam intensity fluctuations.
It is then possible to express the normalized intensity-intensity correlation function modified by fluctuating background as
\begin{equation}
g^{(2)}(x_1, x_2) = \frac{\langle I_0(x_1) I_0(x_2) \rangle + \langle I_B(x_1) I_B(x_2) \rangle + \langle I_0(x_1)\rangle \langle I_B(x_2) \rangle + \langle I_B(x_1)\rangle \langle I_0(x_2) \rangle}{(\langle I_0(x_1)\rangle+\langle I_B(x_1)\rangle)(\langle I_0(x_2) \rangle+\langle I_B(x_2)\rangle)},
\label{app3::g_2}
\end{equation}
where the ensemble average $\langle I_0(x_1) I_0(x_2) \rangle = g_{\Sigma}^{(2)}(x_1,x_2)\cdot\langle I_0(x_1)\rangle\langle I_0(x_2)\rangle$.
The background average intensity and intensity-intensity correlation function are assumed to have the form
\begin{eqnarray}
\langle I_B(x) \rangle & = & C\ ,\\
\langle I_B(x_1) I_B(x_2) \rangle & = & C^2(1+A\delta_{x_1,x_2}) \ ,
\label{app3::backs}
\end{eqnarray}
where $C<<max(I_0)$ and the background signal is therefore not significant in the center of the beam.

\begin{widetext}
	\begin{table}
		\caption{\label{tab::sim_param} Beam parameters used in simulations}
		\begin{center}
			\begin{tabularx}{\textwidth}{l
					>{\centering\let\newline\\\arraybackslash\hspace{0pt}}X |>{\centering\let\newline\\\arraybackslash\hspace{0pt}}X>{\centering\let\newline\\\arraybackslash\hspace{0pt}}X}
				\hline
				\hline
				Simulation                 & \multicolumn{1}{c}{Model I}   & \multicolumn{2}{c}{Model II}   \\
				\hline
				Beam number                    &   1  & 1      & 2         \\
				\hline
				Bandwidth, $2.355\sigma_{\omega}/\omega$ & \multicolumn{1}{c}{$4.4\cdot10^{-5}$}   & \multicolumn{2}{c}{$4.4\cdot10^{-5}$}   \\
				Relative intensity             &  1    & 1      & 0.1        \\
				Beam position, $x_0$ (mm)                                                &        0  & -                 & -      \\
				\hspace{2.5cm} $y_0$ (mm)                                                &        -  & 0                 & -1.5      \\
				Beam size (rms), $\sigma$ (mm)               &  1.6  & 1.4      & 0.5        \\
				Transverse coherence length, $\xi$ (mm)                   & 10   & 8         & 10      \\
				Central frequency, $\omega_0$ (as$^{-1}$)        & 12.6  & 12.6                  & 12.6            \\
				Spectral width, $\Omega$ (fs$^{-1}$)        & 6.3  & 6.3         & 6.3           \\
				Spectral coherence width, $\Omega_c$ (fs$^{-1}$)         & 0.2  & 0.2           & 1.1       \\
				\hline
				\hline
			\end{tabularx}
		\end{center}
	\end{table}
\end{widetext}

The final expression that was used for modeling as it follows from Eqs. (\ref{multiple::general_g} - \eqref{app3::backs}) has the form
\begin{equation}
g^{(2)}(x_1, x_2) = \frac{g_{\Sigma}^{(2)}(x_1,x_2)\langle I_{\Sigma}(x_1)\rangle\langle I_{\Sigma}(x_2)\rangle + C^2(1+A\delta_{x_1,x_2}) + C \langle I_{\Sigma}(x_1)\rangle + C \langle I_{\Sigma}(x_2) \rangle}{(\langle I_{\Sigma}(x_1)\rangle+C)(\langle I_{\Sigma}(x_2) \rangle+C)}.
\label{app3::g_2_full}
\end{equation}

In simulations we used two models (see Table~\ref{tab::sim_param}): one in the horizontal direction consisting of a single beam with the size (r.m.s.) 1.6 mm and coherence length 10 mm and second one in the vertical direction consisting of two beams shifted by 1.5 mm and with the relative intensity of 10\%.
The background level was considered to be 2\% of the total intensity in both cases and parameter $A$ in Eq.\eqref{app3::backs} was taken as $A=0.125$.
All further details of all parameters in both models are listed in Table~\ref{tab::sim_param}.

\section{Appendix V. Spatial degree of coherence and contrast values}

Our experimental results also allowed us to determine the degree of spatial coherence $\zeta_S$ of LCLS radiation for hard x-rays.
Similar to our previous work~\cite{Singer2013, Singer2013_Erratum, Gorobtsov2017}, we obtained this value by applying the following relation
\begin{equation}
\zeta_S = \frac{\int \left| W(\mathbf{r_1}, \mathbf{r_2})\right|^2 d\mathbf{r_1} d\mathbf{r_2}}{\left(\int \langle I(\mathbf{r}) \rangle d\mathbf{r} \right)^2}
= \frac{\int \left| \mu (\mathbf{r_1}, \mathbf{r_2})\right|^2 \langle I(\mathbf{r_1}) \rangle \langle I(\mathbf{r_2}) \rangle d\mathbf{r_1} d\mathbf{r_2}}
        {\left(\int \langle I(\mathbf{r}) \rangle d\mathbf{r} \right)^2}
\label{coherence::coherence_degree_expressed}
\end{equation}
and substituting values of $\left| \mu (\mathbf{r_1}, \mathbf{r_2})\right| $ obtained from the HBT interferometry analysis.
Performing this analysis we determined the spatial degree of coherence for each Bragg peak for both samples see Table~\ref{tab::peak_data}.

Evaluation of the contrast values was performed based on their values determined from the main diagonal of intensity-intensity correlation function $g^{(2)}(x, x)$.
As a final value the mean value of $g^{(2)}(x,x)$ over the region of FWHM of the averaged Bragg peak intensity $\langle I(x) \rangle$ was considered.

\begin{widetext}
	\begin{table}
		\caption{\label{tab::peak_data}Contrast values $\zeta_2$ and spatial degree of coherence $\zeta_S$ determined at different Bragg peaks and crystals}
		\begin{center}
			\begin{tabularx}{\textwidth}{l
					|>{\centering\let\newline\\\arraybackslash\hspace{0pt}}X >{\centering\let\newline\\\arraybackslash\hspace{0pt}}X>{\centering\let\newline\\\arraybackslash\hspace{0pt}}X
					>{\centering\let\newline\\\arraybackslash\hspace{0pt}}X
					|>{\centering\let\newline\\\arraybackslash\hspace{0pt}}X
					>{\centering\let\newline\\\arraybackslash\hspace{0pt}}X
					>{\centering\let\newline\\\arraybackslash\hspace{0pt}}X
					>{\centering\let\newline\\\arraybackslash\hspace{0pt}}X}
				\hline
				\hline
				Crystal                 & \multicolumn{4}{c}{sample 1 (PS 160 nm)}   & \multicolumn{4}{c}{sample 2 (PS 420 nm)}   \\
				\hline
				Peak number & 1 & 2 & 3 & 4 & 1 & 2 & 3 & 4 \\
				\hline				
				
				$\zeta_2$, $x$ & 0.41 & 0.40 & 0.41 & 0.40 & 0.41 & 0.43 & 0.41 & 0.43 \\
				
				$\zeta_2$, $y$ & 0.42 & 0.41 & 0.43 & 0.41 & 0.44 & 0.46 & 0.45 & 0.41 \\
				
				$\zeta_S$, $x$ & 0.92 & 0.94 & 0.94 & 0.93 & 0.93 & 0.89 & 0.95 & 0.88 \\
				
				$\zeta_S$, $y$ & 0.90 & 0.91 & 0.90 & 0.91 & 0.87 & 0.82 & 0.85 & 0.93 \\
				
				\hline
				\hline
			\end{tabularx}
		\end{center}
	\end{table}
\end{widetext}


\section{Appendix VI. Mode analysis and electron bunch energy filtering}

Jitter in the energy of the electron bunch introduces additional problem for the analysis of the monochromator filtered radiation.
If electron bunch energy of a pulse is significantly different from an average, the central wavelength of the pulse is too far removed from the monochromator transmittance band.
\begin{figure}[h]
	\includegraphics[width=10cm]{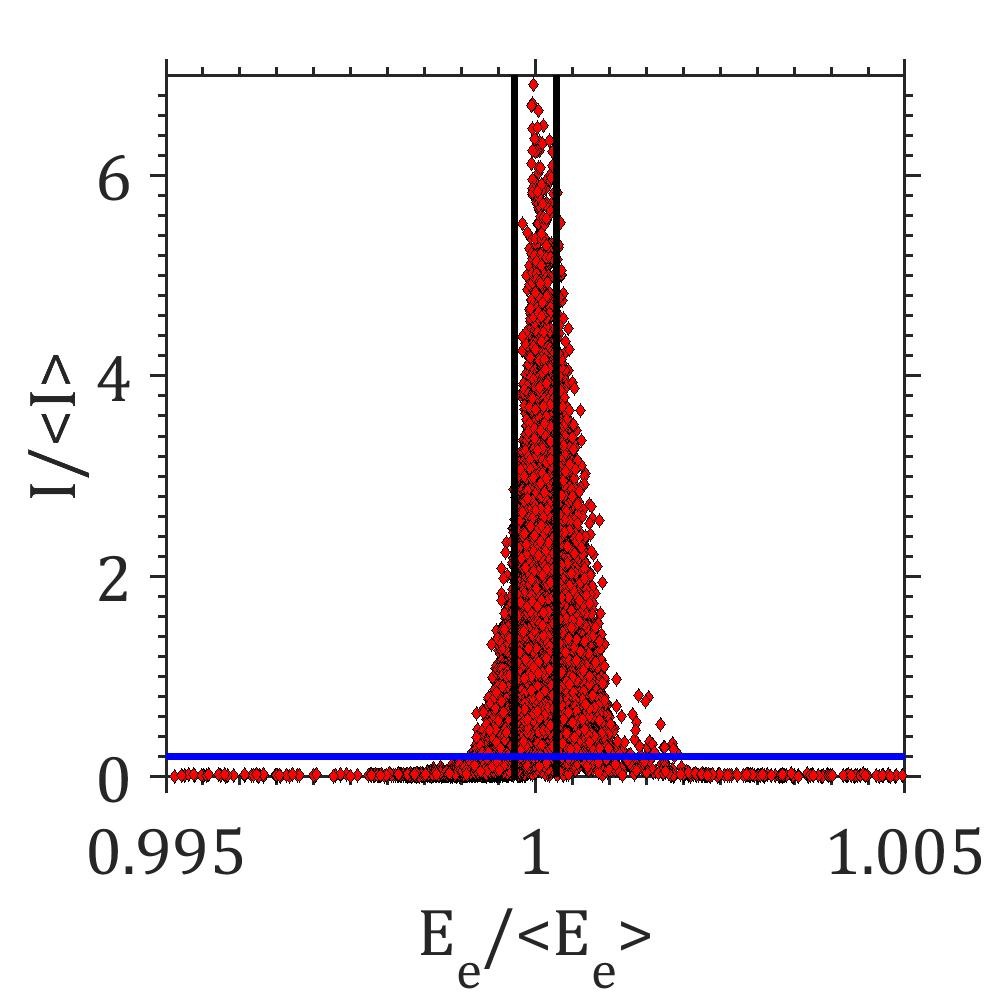}
	\caption{\label{fig::filt}
		Distribution of normalized pulse intensities and corresponding electron bunch energies for the run with the sample 1.
		Black lines show the limits of the filtered region.
		Blue line shows the cutoff used for the histogram fitting without filtering.
	}
	\newpage
\end{figure}

In such a case pulse intensity after monochromator will be significantly reduced, affecting observed statistics.
This is clearly observed in Fig.~\ref{fig::filt}, where the distribution of pulse intensities and corresponding electron bunch energies is shown.

\begin{figure}[h]
	\includegraphics[width=10cm]{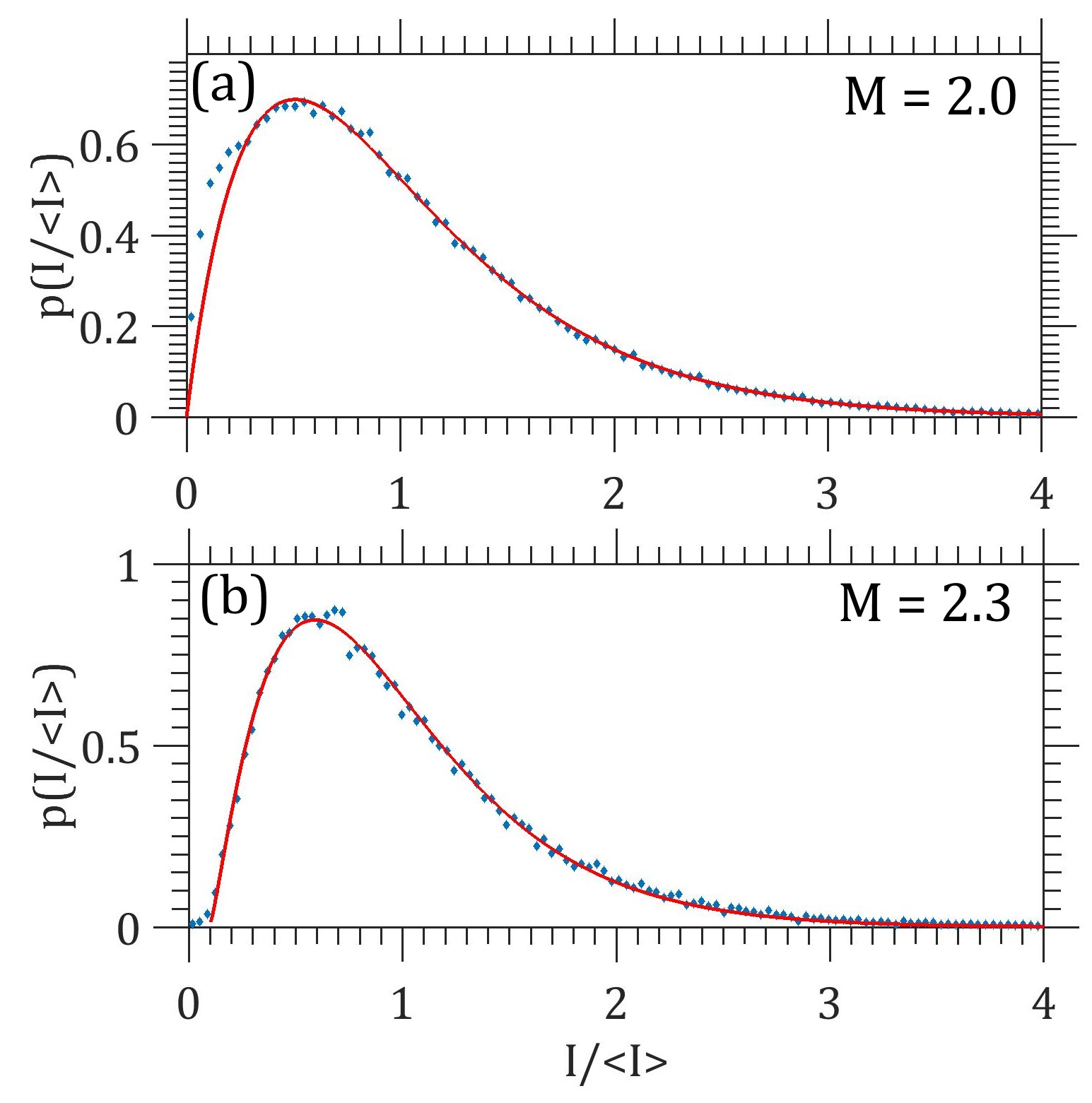}
	\caption{\label{fig::histsupp_0}
		Histogram of the integrated intensity (sample 1) from the Bragg peak 4 in the diffraction pattern before the electron bunch energy filtering (a) and after filtering (b).
	}
	\newpage
\end{figure}

Therefore, it is important to filter the collected pulses by bunch energy.
The filtering was performed by choosing only the pulses for which
\begin{equation}
E_{el}-\langle E_{el} \rangle < \sigma_{E_{el}}/2 \ ,
\label{app::filtering}
\end{equation}
where $E_{el}$ is the electron bunch energy and $\sigma_{E_{el}}^2$ is the energy dispersion (see Fig.~\ref{fig::filt} for the region considered for the following analysis).
Around $50,000$ pulses were left in each run after the filtering.
The difference in the intensity distribution because of filtering can be seen if Fig.~\ref{fig::histsupp_0}, where the histogram of the inegrated intensity from the Bragg peak in the diffraction pattern before and after the electron bunch filtering is shown.
The number of modes is clearly underestimated without filtering.

\begin{figure}[h]
	\includegraphics[width=10cm]{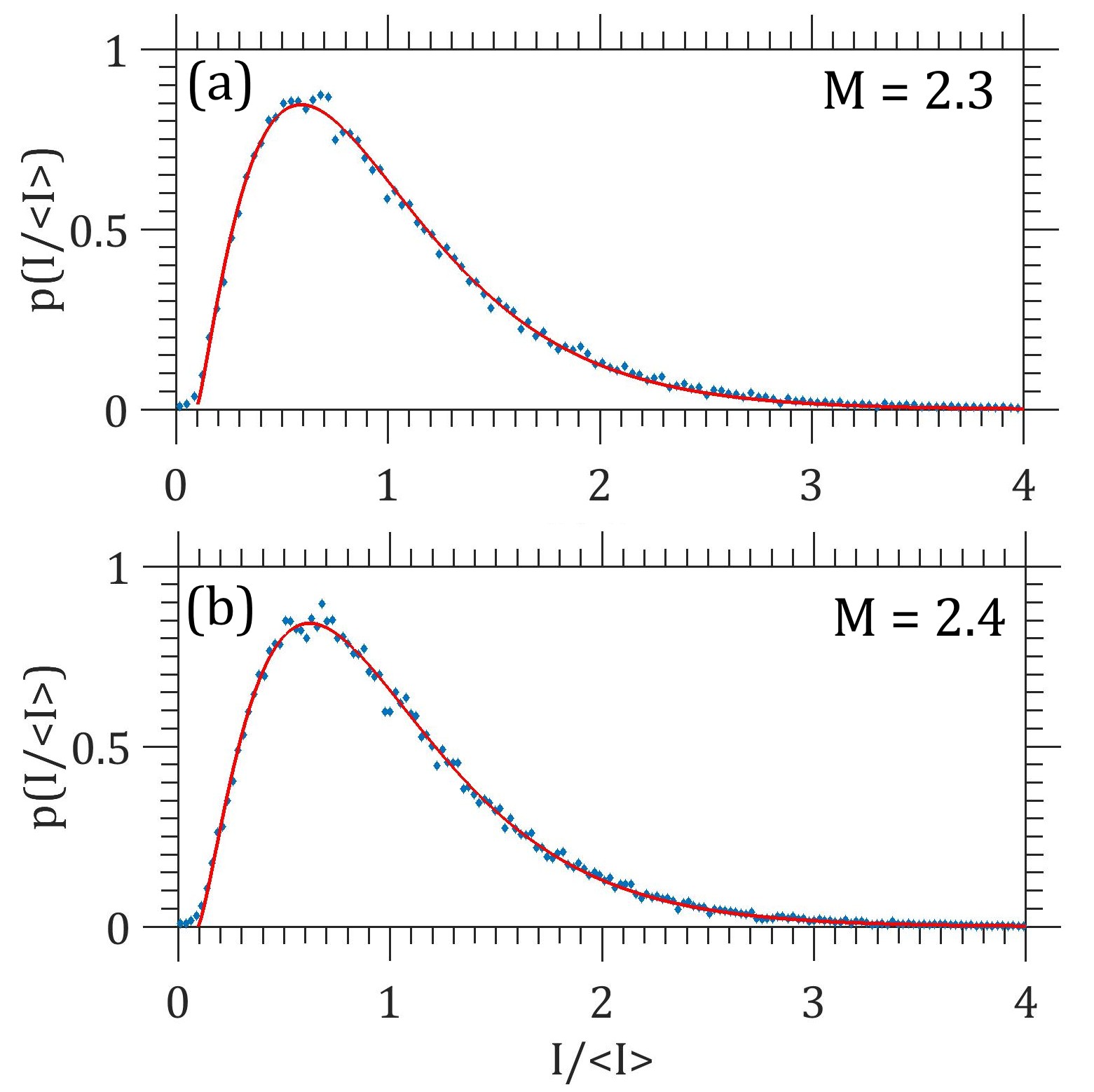}
	\caption{\label{fig::histsupp}
		Histogram of the integrated intensity (sample 1) from the Bragg peak 4 in the diffraction pattern (a) and from the intensity monitor after the monochromator (b).
	}
	\newpage
\end{figure}

The pulse duration was also determined by using the mode analysis of the radiation as suggested in  Ref.~\cite{Saldin1998}.
According to this approach an average number of modes of radiation $M$ is inversely proportional to the normalized dispersion of the energy distribution, that in our case coincide with the contrast function defined in Eq.~\eqref{supp::g2_final_spectr} $M=1/\left[\zeta_2(\sigma_{\omega})\right]$.
Substituting this relation in Eq.~\eqref{coherence::zeta_2_approx1} we obtain for the pulse duration
\begin{equation}
T = \frac{2.355}{2\sigma_{\omega}}\sqrt{M^2-1} \ .
\label{duration_mode}
\end{equation}
We determined the number of modes by fitting integrated intensity distribution at one of the Bragg peaks by Gamma distribution~\cite{Saldin1998} (see Fig.~\ref{fig::histsupp}).
As a result, the number of longitudinal modes was $M \approx 2.3\pm0.1$ and reproducible between different runs.
Substituting this number in Eq.~\eqref{duration_zeta} gives for the pulse duration $11.5 \pm 0.5$ fs.

\end{document}